\documentclass{aastex631}
\usepackage{xcolor}

\newcommand\xmm{XMM-Newton}
\newcommand\swift{Swift}
\newcommand\ixpe{IXPE}
\newcommand\onees{1ES~0229+200}

\def\arcsec{{\rm\thinspace arcsec}}
\def\cm{{\rm\thinspace cm}}

\def\erg{{\rm\thinspace erg}}

\def\g{{\rm\thinspace g}}

\def\G{{\rm\thinspace G}}

\def\K{{\rm\thinspace K}}
\def\keV{{\rm\thinspace keV}}

\def\m{{\rm\thinspace m}}

\def\s{{\rm\thinspace s}}

\def\ergpcmsqps{\hbox{$\erg\cm^{-2}\s^{-1}\,$}}

\shorttitle{X-ray Polarization of \onees}
\shortauthors{Ehlert et al.}
\graphicspath{{./}{figures/}}

\begin{document}

\title{X-ray Polarization of the BL Lac Type Blazar \onees}

\author[0000-0003-4420-2838]{Steven R. Ehlert}
\correspondingauthor{Steven R. Ehlert}
\email{steven.r.ehlert@nasa.gov}
\affiliation{NASA Marshall Space Flight Center, Huntsville, AL 35812, USA}
\author[0000-0001-9200-4006]{Ioannis Liodakis}
\affiliation{NASA Marshall Space Flight Center, Huntsville, AL 35812, USA}
\affiliation{Finnish Centre for Astronomy with ESO,  20014 University of Turku, Finland}
\author[0000-0001-9815-9092 ]{Riccardo Middei}
\affiliation{Space Science Data Center, Agenzia Spaziale Italiana, Via del Politecnico snc, 00133 Roma, Italy}
\affiliation{INAF Osservatorio Astronomico di Roma, Via Frascati 33, 00078 Monte Porzio Catone (RM), Italy}
\author[0000-0001-7396-3332]{Alan P. Marscher}
\affiliation{Institute for Astrophysical Research, Boston University, 725 Commonwealth Avenue, Boston, MA 02215, USA}
\author[0000-0003-0256-0995]{Fabrizio Tavecchio}
\affiliation{INAF Osservatorio Astronomico di Brera, Via E. Bianchi 46, 23807 Merate (LC), Italy}

\author[0000-0002-3777-6182]{Iván Agudo}
\affiliation{Instituto de Astrof\'{i}sica de Andaluc\'{i}a, IAA-CSIC, Glorieta de la Astronom\'{i}a s/n, 18008 Granada, Spain}
\author[0000-0002-9328-2750]{Pouya M. Kouch}
\affiliation{Finnish Centre for Astronomy with ESO, 20014 University of Turku, Finland}
\affiliation{Department of Physics and Astronomy, 20014 University of Turku, Finland}

\author[0000-0002-9155-6199]{Elina Lindfors}
\affiliation{Finnish Centre for Astronomy with ESO, 20014 University of Turku, Finland}

\author[0000-0002-1445-8683]{Kari Nilsson}
\affiliation{Finnish Centre for Astronomy with ESO, 20014 University of Turku, Finland}

\author[0000-0003-3025-9497]{Ioannis Myserlis}
\affiliation{Institut de Radioastronomie Millim\'{e}trique, Avenida Divina Pastora, 7, Local 20, E–18012 Granada, Spain}
\affiliation{Max-Planck-Institut f\"{u}r Radioastronomie, Auf dem H\"{u}gel 69,
D-53121 Bonn, Germany}

% ...




\begin{abstract}
We present polarization measurements in the $2–8 \keV$ band from blazar \onees, the first extreme high synchrotron peaked source to be observed by the Imaging X-ray Polarimetry Explorer (\ixpe). Combining two exposures separated by about two weeks, we find the degree of polarization to be $\Pi_{X} = 17.9 \pm 2.8 \%$ at an electric-vector position angle $\psi_X = 25.0 \pm 4.6^{\circ}$ using a spectropolarimetric fit from joint \ixpe{} and \xmm{} observations. There is no evidence for the polarization degree or angle varying significantly with energy or time on both short time scales (hours) or longer time scales (days). The contemporaneous polarization degree at optical wavelengths was $>$7$\times$ lower, making \onees{} the most strongly chromatic blazar yet observed. This high X-ray polarization compared to the optical provides further support that X-ray emission in high-peaked blazars originates in shock-accelerated, energy-stratified electron populations, but is in tension with many recent modeling efforts attempting to reproduce the spectral energy distribution of \onees{} which attribute the extremely high energy synchrotron and Compton peaks to Fermi acceleration in the vicinity of strongly turbulent magnetic fields. 
\end{abstract}



\section{Introduction} \label{sec:intro}

The jets of active galactic nuclei (AGN) have been shown to be luminous sources at energies spanning from the lowest frequency radio waves to the highest energy gamma rays. These observations clearly show that AGN are powerful sites of non-thermal radiation originating from the acceleration of particles to highly relativistic energies. One particularly important subclass of AGN for studying the mechanisms of particle acceleration and their subsequent radiation are blazars. Blazars are AGN whose relativistic, highly energetic plasma jets are pointed toward our line of sight \citep[e.g.,][]{Blandford2019,Hovatta2019}. Blazar emission is usually characterized by two broad spectral humps from radio to X-rays and X-rays to TeV $\gamma$-rays. The low-energy hump is interpreted as synchrotron radiation from energetic electrons, which for the most extreme blazars can peak at energies of $\sim 1 - 50 \keV$
\citep{Costamante2001,Liodakis2022,DiGesu2022}. 

One of the most extreme blazars, at least from the standpoint of its panchromatic spectral energy distribution (SED) observed to date is \onees{}  ($\rm RA=02h32m48.6s$, Dec=+20$\rm^o$17$^\prime$17.4$^{\prime\prime}$, $z=0.14$). \onees{} is a BL Lacertae object which belongs to the rare class of extreme high synchrotron peaked (HSP) objects \citep{Costamante2002, Biteau2020} with a synchrotron peak frequency of $\rm \nu_{syn}\sim10^{19}~Hz$ or $\sim 40 
\keV$ \citep{Ajello2020}. Due to its extreme spectrum, 1ES~0229+200 has been used to study the extragalactic background light \citep{Aharonian2007}, extragalactic magnetic fields \citep{Tavecchio2010,Acciari2023}, and Lorentz invariance violations \citep{Tavecchio2016}. The origin of the second emission hump is Compton scattering of photons by higher energy particles, but questions about the origin of the particles responsible for the scattering and the source of the seed photons remain unanswered. Recent results from the Imaging X-ray Polarimetry Explorer (\ixpe ) collaboration for blazars and AGN with synchrotron peaks well below the \ixpe{} bandpass \citep[e.g.,][]{Ehlert2022CenA,Middei2023,Peirson2023} suggest relativistic electrons are the dominant scattering particles, but are not yet sensitive enough to distinguish between different seed photon populations. 

The nature of the extreme HSP sources and the physical processes that lead to such high $\rm \nu_{syn}$ are still unknown. Previous attempts to model the extremely high energy synchrotron and Compton peaks \citep[at $\sim 9-20 \keV$ and $\sim 12 \thinspace \mathrm{TeV}$, respectively;][]{Ajello2020,Costamante2018} in conjunction with single-zone synchrotron self-Compton (SSC) models have resulted in extreme, finely-tuned parameters \citep[e.g.][]{Kaufmann2011}. In particular, the simplest single-zone SSC models predict that \onees{} has the following unusual properties: 1) an electron energy distribution that spans an unusually small range of energies ; 2) a very high jet bulk Doppler factor of $\delta \sim 40$ or larger;  and 3) extremely weak magnetic fields with $B \sim 30 \thinspace \mathrm{\mu G}- 2 \thinspace \mathrm{m G} $ \citep{Kaufmann2011,Costamante2018}. Models with these parameters are also dramatically out of equipartition, with electron energy densities orders of magnitude higher than the magnetic field energy density. For these reasons, more sophisticated particle acceleration models for these extreme HSP sources have been proposed \citep[e.g.][]{Zech2021,AguilarRuiz2022,Tavecchio2022}. Although the details of these different proposed models can differ significantly, they all result in fits that reproduce the panchromatic SED of \onees{} with somewhat stronger magnetic fields ($B \sim 10^{-3} \thinspace \mathrm{G}$). The X-ray synchrotron photons in these models originate from electrons that have recently undergone stochastic acceleration (SA) and/or diffuse shock acceleration (DSA) in the turbulent, magnetized plasma of the jet.

The IXPE satellite, launched in December 2021 \citep{Weisskopf2022}, is the first X-ray polarization mission, offering a new way to study high energy and extreme phenomena in the Universe. It is particularly important for the study of extragalactic jets, as it can directly test particle acceleration and emission mechanisms in blazars by providing information on the geometry of the magnetic fields involved \citep{Zhang2013,Zhang2016,Liodakis2019,Peirson2022}. During the first year (2022) of IXPE, several observations of Mrk~501 \citep{Liodakis2022} and Mrk~421 \citep{DiGesu2022} were obtained, with more HSPs scheduled to be observed in the second year.  Here we present the first X-ray polarization observations of \onees . All of the prior X-ray and multiwavelength polarization observations of HSPs point to a shock-accelerated, energy stratified electron population for the origin of the synchrotron X-rays in blazar jets \citep{Marscher1985,Angelakis2016,Tavecchio2021}. However, this is the first time IXPE has observed an extreme HSP. In Section \ref{sec:xray_obs} and Section \ref{sec:xray_modeling} we present our X-ray observations and modeling, in Section \ref{sec:xray_pol} and Section \ref{sec:xray_pol_var} the X-ray polarization results, and in Section \ref{sec:multi_pol} our contemporaneous radio and optical observations.
In Section \ref{sec:disc_conc} we discuss our results and present our conclusions.
Unless otherwise noted, all uncertainties described and error bars plotted correspond to 68.3\% (1-$\sigma$) confidence intervals for the measurement in question.

\section{X-ray Observations}\label{sec:xray_obs}

\subsection{\ixpe}
\ixpe{} is a NASA mission in partnership with the Italian Space Agency (ASI). As described in detail in \citet[][and references therein]{Weisskopf2022}, the \ixpe{} observatory includes three identical X-ray telescopes, each comprising an X-ray mirror assembly (NASA-furnished) and a polarization-sensitive pixelated detector (ASI-furnished), to provide imaging polarimetry over a nominal 2-8 $\keV$ band.  IXPE data telemetered to ground stations in Malindi (primary) and in Singapore (secondary) are transmitted to the Mission Operations Center (MOC, at the Laboratory for Atmospheric and Space Physics, University of Colorado) and then to the Science Operations Center (SOC, at NASA Marshall Space Flight Center). Using software developed jointly by ASI and NASA, the SOC processes science and relevant engineering and ancillary data, to produce data products that are archived at the High-Energy Astrophysics Science Archive Research Center (HEASARC, at NASA Goddard Space Flight Center), for use by the international astrophysics community.
 
 \ixpe{} targeted 1ES~0229+200 starting on 2023 January 15 with its three Detector Units (DU's). \ixpe{} observations were taken in two separate segments. The first exposure took place from UT 2023-01-15T10:02 to 2023-01-18T16:44, while the second occurred from UT 2023-01-27T00:53 to 2023-02-01T14:44. A total of 401 ks of total exposure on source was collected, with 37\% and 63\% of the total exposure time taking place in the first and second time segments, respectively.  Cleaned level 2 event files were computed and calibrated using standard filtering criteria with the dedicated \textsc{ftools} tasks and latest \ixpe~ calibration data base (version 20211118). Likely background events were removed from the event lists using the selection criteria of \cite{DiMarco2023}. Stokes $Q$ and $U$ background spectra were derived from source-free circular regions with a radius of 102 \arcsec . Extraction radii for the I Stokes spectra of \onees{} were selected via an iterative process aimed at maximizing the signal-to-noise ratio (SNR) in the 2–8 keV energy band. This method is similar to the approach described in \citet{Piconcelli2004}. We thus adopted circular regions centred on the target with radii of 42 \arcsec\ for DU1 and 47 \arcsec\ for both DU2 and DU3. \footnote{DU1 has a slightly sharper PSF than DU2 and DU3.} The same radii were also used for the $Q$ and $U$ Stokes spectra. We then binned the Stokes I spectra, requiring an SNR higher than 5 in each spectral channel, while a constant binning (0.2 keV) was adopted for the $Q$ and $U$ Stokes spectra.

\subsection{\xmm}
 \xmm~ \citep{Jansen2001} observed 1ES~0229+200 on 2023 January 15, quasi-simultaneously with IXPE, for about 18 ks. We extracted
the event lists of the European Photon Imaging Camera \citep[EPIC-pn;][]{Struder2001} with the standard System Analysis Software (SAS, version 19.0.0) and the  calibration database corresponding to this release. We extracted the source spectrum by selecting a region in the CCD image of 40 \arcsec\ radius centered on the source, and the background by extracting a source-free larger region (radius=50 \arcsec). The response matrices and auxiliary response files were generated with the SAS commands {\it rmfgen} and {\it arfgen}, respectively. Spectra were then grouped by allowing 30 counts for each spectral bin in order not to over-sample the instrumental resolution by a factor larger than 3.

\subsection{\swift}

Before, during and after the \ixpe \ pointing, we monitored 1ES~0229+200 with the Neil Gehrels Swift X-Ray Telescope (Swift-XRT).
The Swift-XRT observations each had an exposure time of $\sim1$ ks and were performed in Photon Counting (PC) mode. We used the XRT Data Analysis Software
(XRTDAS\footnote{\url{https://swift.gsfc.nasa.gov/analysis/xrt_swguide_v1_2.pdf}} , v. 3.6.1) to reduce, clean and process the data. In the analysis, we used the latest calibration files available in the  Swift-XRT CALDB (version 20220331). The source spectrum was extracted from the cleaned event file, adopting a circular region with a radius of 47 \arcsec . A concentric annulus with inner (outer) radii of 120 (150) \arcsec\ was then adopted to determine the background. The background was computed using long exposures available in the Swift archive. Finally, spectra were binned to achieve at least 25 counts in each energy bin.\\

\section{Spectral Modeling}\label{sec:xray_modeling}

We fit the joint \xmm{} and \ixpe{}  Stokes $I$ spectral data to an absorbed log-parabolic model of the form \texttt{const * tbabs * logpar} within XSPEC. The log-parabola model is a straightforward extension of a simple power-law model, and parameterized as

\begin{equation}
    \frac{dN}{dE} = K \left(\frac{E}{E_{\mathrm{piv}}}\right)^{-\Gamma + \beta \log(E/E_{\mathrm{piv}})} .
\end{equation} 
This model is commonly fit to X-ray spectra of blazars that show evidence of curvature beyond a single power-law. We find best-fit parameters of $\Gamma = 1.82 \pm 0.01$ and $\beta = 0.13 \pm 0.02$, indicating that the spectrum is steepening with increasing energy. The Galactic absorption column density was frozen at $N_\mathrm{{H}} = 7.81 \times 10^{20} \cm^{-2}$, as determined by the HI4PI survey \citep{HI4PI2016} for our fiducial model, and the pivot energy was fixed to $E_{piv} = 1\keV$. Allowing the column density to be freely fit by the model results in a best-fit value of $N_\mathrm{{H}} = 8.01 \pm 0.01 \times 10^{20} \cm^{-2}$ and negligible improvement to the overall fit.    

\begin{figure}
    \centering
    \includegraphics[width=0.8\textwidth]{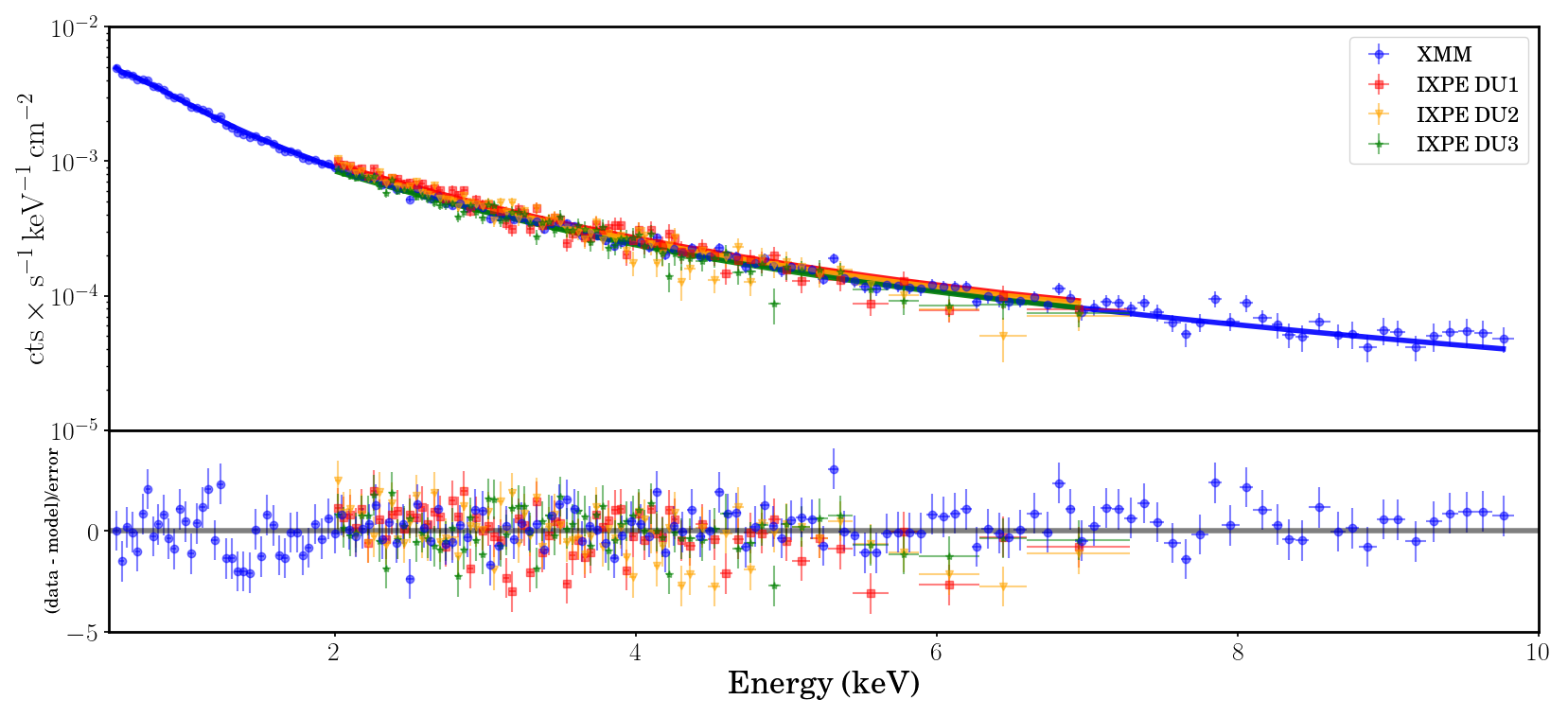}
    \caption{The best-fit spectral model to the \xmm{} and \ixpe{} Stokes I data for \onees. \textit{Top}: The spectra with the best-fit log-parabolic model overlaid, corresponding to $\Gamma = 1.82 \pm 0.01$ and $\beta = 0.13 \pm 0.02$. \textit{Bottom}: The normalized residuals of the data with respect to the best-fit model.}
    \label{fig:ispectra}
\end{figure}

The overall $\chi^2$ value of this fit is $\chi^{2} = 409$ with 370 degrees of freedom. We compare these values to a simple power-law model, which has a best-fit photon index value of $\Gamma = 1.894 \pm 0.006$ with $\chi^2 = 465$ and 371 degrees of freedom. When $E_{\mathrm{piv}} = 1 \keV$, the log-parabola model is a simple extension of a power-law model, and we can use an F-test to compare the statistical significance of the additional parameter to the fit improvement. Under a null hypothesis where the true model is a power-law, the probability of such an improvement is $P_{\mathrm{null}} \sim 6 \times 10^{-12}$.  We can also use the Akaike and Bayesian Information Criteria to compare their overall fit quality. For this total change in the fit statistic ($\Delta \chi^{2} = -56.12$), both criteria strongly favor the log-parabolic model over the simple power-law model after accounting for the additional parameters ($\Delta AIC = -52$, $\Delta BIC = -52$).  We therefore consider this log-parabolic model our fiducial spectrum for the remainder of this work. The \texttt{const} term, which accounts for cross-calibration terms in the effective areas of the four different detectors, was fixed to unity for \ixpe{} DU1. The best-fit constant normalization offsets for DU2, DU3, and our \xmm{} spectra are $0.96 \pm 0.02$, $0.89 \pm 0.01$, and $0.92 \pm 0.01$, respectively. These factors are consistent with previous results reported by spectral fits using \ixpe{} \citep{[e.g.][]Ehlert2022CenA}. The spectra from all four detectors (\ixpe{} DU1, \ixpe{} DU2, \ixpe{} DU3, and \xmm), along with the best-fit log-parabolic model, are displayed in Figure \ref{fig:ispectra}. 

Although the log-parabola model deployed here represents a ``good'' fit to the data, the best-fit parameters differ significantly from the results of \cite{Costamante2018}, who used an identical model to fit \swift{} and \textit{NuSTAR} observations of \onees{}. In their best-fit model,\footnote{We note that \cite{Costamante2018} also used the same $1 \keV$ pivot energy as this work.} $\Gamma = 1.49 \pm 0.04$ and $\beta = 0.27 \pm 0.02$. These different values suggest that \onees{} during the \ixpe{} observations had a much softer photon index at $1 \keV$, but did not further soften as rapidly with energy as it did in 2013 when the \textit{NuSTAR} observations were performed.  During our observations, the X-ray flux is at the $\rm1\times10^{-11} \ergpcmsqps$ level in the 2-10~keV band, whereas during the \cite{Costamante2018} observations it is about a factor of two brighter ($\rm1.95\times10^{-11} \ergpcmsqps$). 1ES0229+200 becomes harder-when-brighter as was found in \cite{Acciari2020}, hence the difference in spectral shape could potentially arise from the difference in flux. However, comparing these results directly is subject to the caveat that our analysis considers a much lower energy band than the result of \cite{Costamante2018}

\begin{figure}
    \centering
    \includegraphics[width=0.95\textwidth]{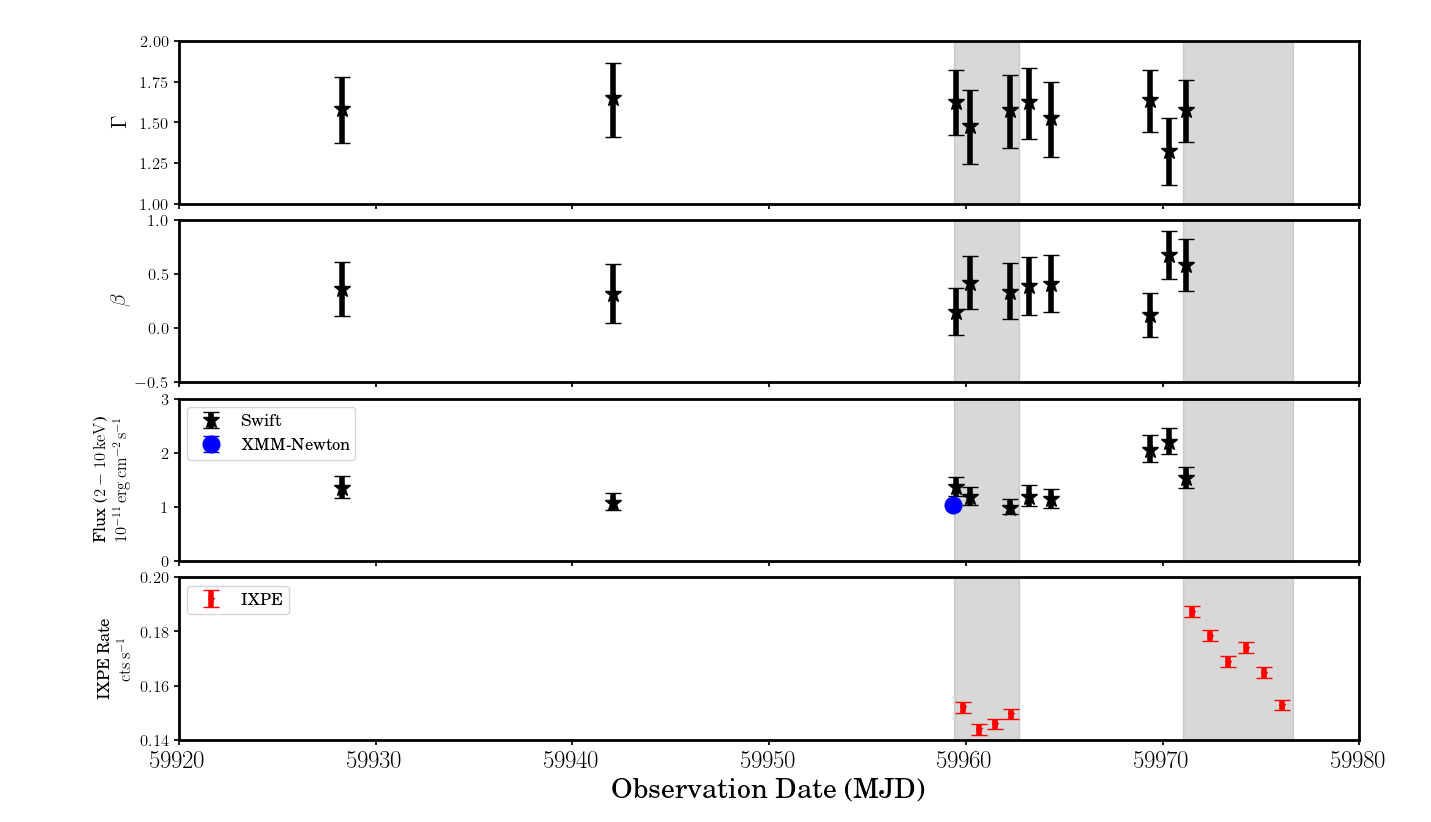}
    \caption{The best-fit spectral model parameters and flux of \onees{} as determined by \swift{} observations taken before and during the \ixpe{} observations. The shaded gray regions correspond to the two segments where \ixpe{} was observing \onees. \textit{Top Panel:} The best-fit values for $\Gamma$, the photon index at $1 \keV$, as a function of time. \textit{Upper-Middle Panel: } The best-fit values for $\beta$, the curvature parameter, as a function of time. \textit{Lower-Middle Panel: } The \swift{} 2-10 $\keV$ flux as a function of time, in units of $10^{-11} \ergpcmsqps$. We have also included the corresponding flux measurement from the \xmm{} observation in blue, which is consistent with the measurements from \swift. \textit{Bottom Panel: } The \ixpe{} count rate during the two segments. The parameters for each of the fits to the \swift{} data are consistent with each other, but slightly softer than the measurements of \cite{Costamante2018} when \onees{} was a factor of two brighter.} 
    \label{fig:swiftfits}
\end{figure}

We present best-fit log-parabolic models for each of the individual \swift{} observations of \onees{} along with their fluxes in the 2-10 $\keV$ band in Figure \ref{fig:swiftfits}. This figure also includes the time-resolved \ixpe{} light curve for each observation segment. Each of these spectral fits was performed over the 0.3-10 $\keV$ energy band. 
Both the \swift{} and \ixpe{} data show evidence that \onees{} became brighter during the second \ixpe{} observation segment. We have fit the same log-parabolic model to each \ixpe{} observation segment to test for changes in the spectral parameters. No changes in the spectral parameters between the two segments were identified beyond the overall normalization of the model. The values of $\Gamma$ for the first and second segments are $\Gamma = 1.69 \pm 0.28$ and $\Gamma = 1.68 \pm 0.19$, respectively. The best-fit values of $\beta$  for the first and second segments are $\beta = 0.41 \pm 0.28$ and $\beta = 0.55 \pm 0.19$, respectively. The normalizations for the second segment are $\sim 2 \times$ larger than for the first segment.   We note that there is a small mismatch between the spectral parameters derived from \swift+\ixpe{} and the \xmm+\ixpe{} fit. This could be due to the observations not being strictly simultaneous, as well as the fact that  \xmm's significantly higher effective area above $\sim 5 \keV$ can better constrain the shape of the spectrum in the $\sim 5-10 \keV$ band. Nevertheless, the parameters are within $1\sigma$ and the polarimetric results (see section \ref{sec:xray_pol}) are not sensitive to small variations of the spectral shape.


\section{Polarization Measurements}\label{sec:xray_pol}

We have determined the broad-band polarization of \onees{} using two different analysis methods: one by measuring the average normalized Stokes parameters over various energy bands to determine the average polarization degree and angle using \textsc{ixpeobssim} \citep{PesceRollins2019,Baldini2022}, and the other by a simultaneous spectro-polarimetric fit of the Stokes $Q$ and $U$ spectra along with the \xmm \ and \ixpe \ Stokes-I spectra. We discuss the results of these two analysis methods separately, since they measure slightly different, albeit related, polarization quantities.

\subsection{Polarization Cube}

Our polarization analysis determines the average polarization using the statistical framework of \cite{Kislat2015} without any event-specific weights. 
Over the entire nominal \ixpe \ bandpass of 2-8 $\keV$ with 33502 net counts, the average values of the Stokes parameters are $Q= 0.063 \pm 0.035$ and $U= 0.110 \pm 0.035$, respectively. Under a 
null hypothesis of zero true polarization, the $\chi^{2}$ for these Stokes parameter values is 12.80 with two degrees of freedom. This value corresponds to a confidence level of 99.8
\%, indicating detection. The corresponding polarization degree of this measurement is $\Pi_{X} = 12.5 \% \pm 3.2 \%$ and the electric-vector polarization angle is $\psi = 30.0\degr \pm 8.0\degr$, east of north. The significance of the detection depends strongly on the choice of energy band. At lower energies of $2-4 \keV$ with 29561 net counts, the normalized Stokes parameter differ from $0$ at the $99.94\%$ confidence level. On the other hand, the 3941 net counts in the $4-8 \keV$ band show no statistically significant evidence of polarization at $99\%$ confidence. The $99\%$ confidence upper limit in the $4-8 \keV$ band is $\Pi_{X} < 29.8\%$. Although no statistically significant polarization is detected in the $4-8 \keV$ band, the upper limit is consistent with the polarization degree observed at lower energies. The significance of the detection is highest in the $2-6 \keV$ energy band, for which the significance of the detection is securely above $99.99\%$ confidence: $\Pi_{X} = 14.9 \%  \pm 3.0 \%$ and $\psi = 33.9 \degr \pm 5.7 \degr$. 

\begin{figure}
    \centering
    \includegraphics[width=0.6\textwidth]{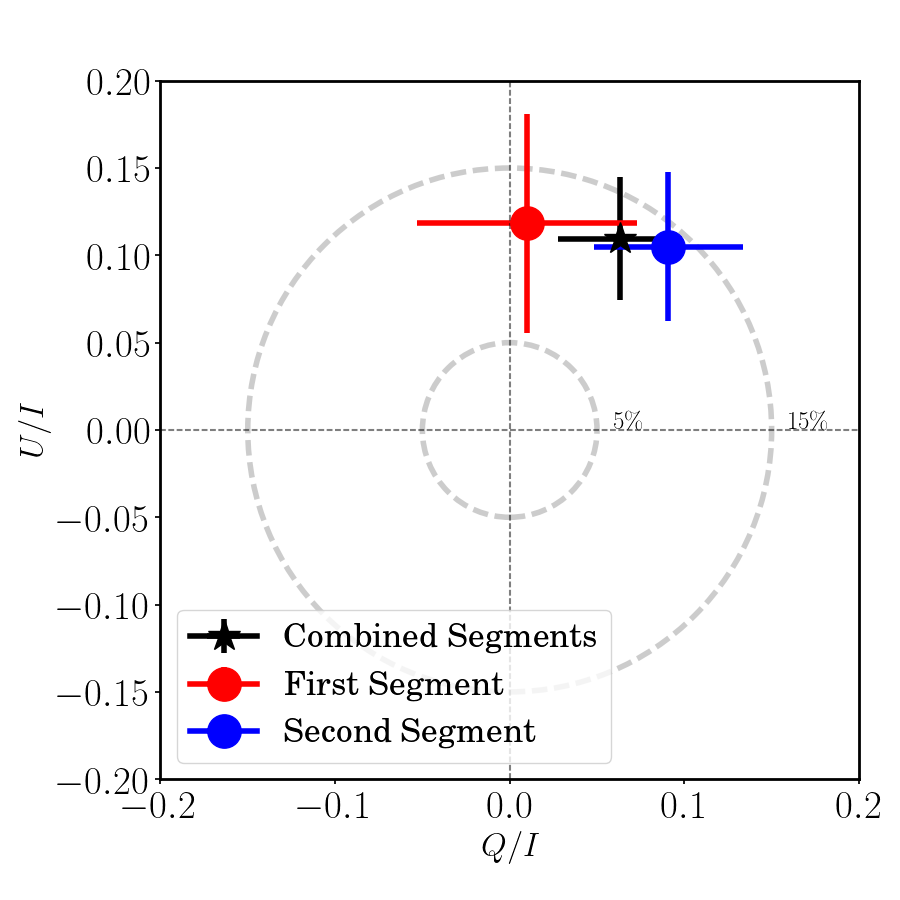}
    \caption{Time-averaged Stokes parameters in the 2-8 $\keV$ band for the two observation segments of \onees. See the main text for  the exposure times and date ranges for each segment. It is clear that the mean Stokes parameters from the first segment (in red) are consistent with those from the second segment (in blue). We can therefore safely combine the events from both segments, the results of which are shown in black. For reference, silver circles with constant polarization degrees of $\Pi = 5\%$ and $\Pi = 15\%$ are also drawn.  }
    \label{fig:qusegments}
\end{figure}

The polarization behavior of this source is consistent between the two observation segments. Restricting the data to only include events gathered during each of the individual segments gives nearly identical results (see Table \ref{tab:qu_segments}). We therefore conclude that we can combine the results from both time segments without loss of any polarization information. The Stokes parameters for each segment, as well as the time-integrated average, are shown in Figure \ref{fig:qusegments}.

\begin{table}[]
    \centering
    \begin{tabular}{c  c  c  c}
           Stokes Parameter & First Segment & Second Segment & Combined\\
         \hline 
           $Q/I$ & $1.0\% \pm 6.3\% $ & $9.1\% \pm 4.3\%$ & $6.3\% \pm 3.5 \%$\\
           $U/I$ & $11.9\% \pm 6.3\%$ & $10.5\%  \pm 4.3\%$ & $11.0 \pm 3.5\%$\\ 
           \hline \hline
    \end{tabular}
    \caption{Normalized Stokes parameters in the 2-8 $\keV$ bandpass for the two \ixpe{} observation segments of \onees{}, as determined by the polarization cube analysis.  }
    \label{tab:qu_segments}
\end{table}

\subsection{Spectro-polarimetric Fits}

To further investigate the extent to which we can detect and measure polarization in \onees, we have added to our spectral fit the $Q$ and $U$ spectra for all three IXPE DU's. We have performed the spectro-polarimetric model fit using \textsc{XSPEC} \citep{Strohmayer2017} by including an energy independent polarization model component to our fiducial log-parabolic model. Unlike the polarization cube analysis, the spectra used for these fits were weighted using the method of \cite{DiMarco2022}. In \textsc{XSPEC} terms, this corresponds to a model of the form \texttt{const * tbabs * polconst * logpar}.  Our best-fit model for the polarization degree and angle from this model is $\Pi_{X} = 17.9 \pm 2.8 \%$ and $\psi = 25.0 \pm 4.6^{\circ}$. The total $\chi^{2}$ of this fit is 586.56 with 542 degrees of freedom. 

\begin{figure}
    \centering
    \includegraphics[width=0.8\textwidth]{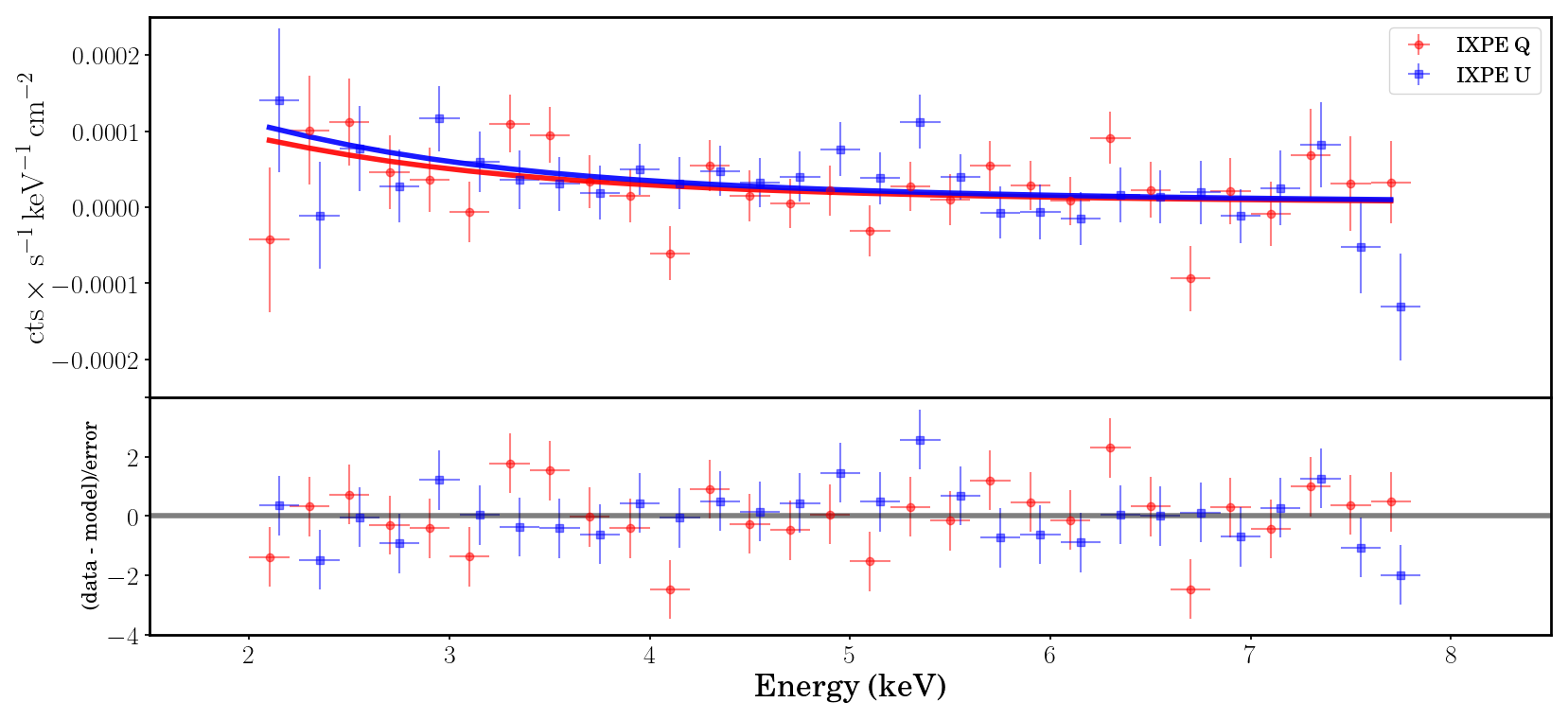}
    \caption{\textit{Top: } Stokes $Q$ and $U$ spectra as measured by \ixpe. For presentation purposes only, we have grouped the spectra from all three \ixpe{} detectors together and shifted the Stokes $U$ spectra by $0.05 \keV$. The solid red and blue curves correspond to the best fit models for the $Q$ and $U$ spectra, respectively. \textit{Bottom:} The residuals between the spectra and the best-fit model as a function of energy.   }
    \label{fig:quspectra}
\end{figure}

The spectro-polarimetric fit enables another hypothesis test for the presence of polarization. Assuming a polarization degree of $\Pi =0$ and fixing $\Psi =0$ results in a fit with $\chi^{2} = 626.36$ with 544 degrees of freedom. The probability of obtaining a $\chi^2$ value equal to or exceeding this under the assumption of zero polarization is $P_{\mathrm{null}} = 0.0028$, providing strong evidence that these data are an improbable realization of the model assuming zero polarization. Furthermore, allowing the polarization degree and angle to be free parameters provides a statistically significant improvement to the overall fit ($\Delta \chi^{2} = -39.8$ with two fewer degrees of freedom), which for these data corresponds to a null hypothesis improvement (as determined by an $F$-test) of $P_{\mathrm{null}} = 1.9 \times 10^{-8}$. The corresponding changes in the Akaike and Bayesian Information Criteria are $\Delta AIC = -36$ and $\Delta BIC = -27$, respectively. The mean Stokes $Q $ and $U$ spectra, along with the best-fit specto-polarimetric model, are shown in Figure \ref{fig:quspectra}. 

\section{Tests for Variability in Polarization Degree and Angle}\label{sec:xray_pol_var}

\subsection{Energy Dependent Polarization}

The non-detection of polarization in the $4-8 \keV$ band, as compared to detection when the energy range is extended down to $2 \keV$, leads to the question of whether this is caused by insufficient photon statistics in the $4-8 \keV$ band or by an energy-dependent polarization. While it is clear that the vast majority of the photons are observed at lower energies, we further test the null hypothesis of constant polarization as a function of energy using our spectro-polarimetric model. We replace the constant polarization model component with a constant + linear energy dependence of the polarization degree and angle (\texttt{const * tbabs * pollin * logpar} in \textsc{XSPEC}). We find that the $90\%$ confidence interval for the polarization degree's linear term is consistent with zero, and that the total improvement in $\chi^{2}$ with respect to the constant polarization model is $\Delta \chi^{2} = -1.1$ for 2 fewer degrees of freedom, entirely consistent with the expected improvement to the fit arising from arbitrary additional parameters. Both of these calculations indicate that the evidence for any dependence of the polarization on energy is marginal. We therefore conclude that a constant polarization degree across the entire $2-8 \keV$ band is consistent with the observations.

\subsection{Time-Dependent Variation in the Stokes Parameters}

\begin{figure}
    \centering
    \includegraphics[width=0.6\textwidth]{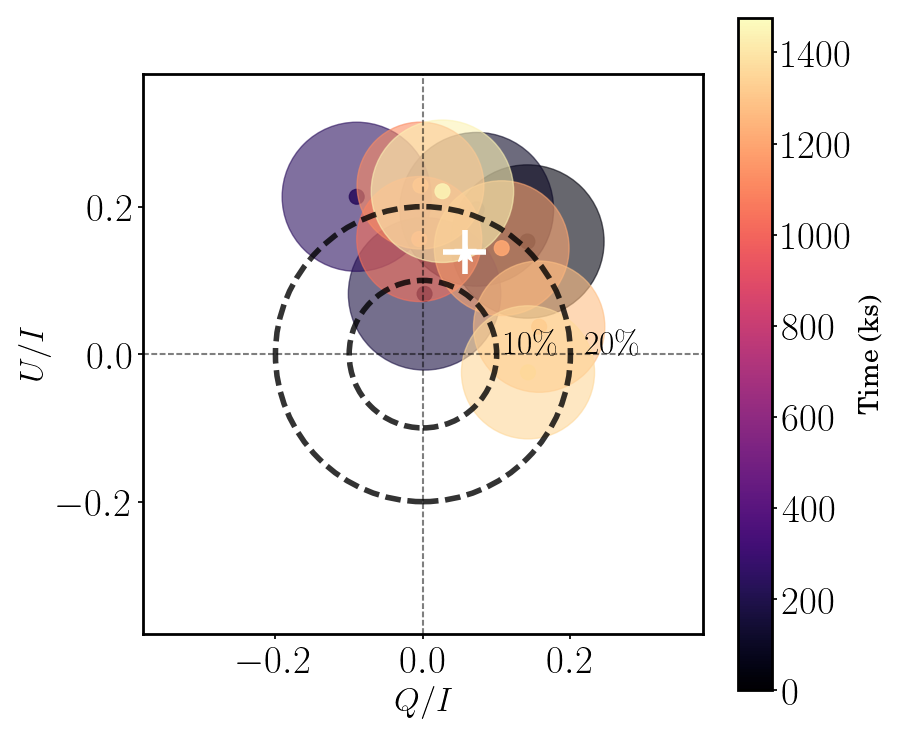}
    \caption{Variations of the Stokes parameters as a function of time during the observations of \onees. The circles correspond to the mean Stokes parameters and their statistical uncertainties in the 2-6 $\keV$ band, with their colors corresponding to the time in ks
    since the beginning of the first observation segment. Both segments are included. The white cross denotes the time-averaged mean Stokes parameters and their uncertainties. As we describe in detail in the text, we are unable to reject a null hypothesis that the true Stokes parameters in each time bin are equal to the time-averaged values.  }
    \label{fig:variability}
\end{figure}

Recent observations of other blazars, in particular Mrk 421 \citep{DiGesu2023} show clear evidence of intra-observation variability. In the case of Mrk 421, this variability is consistent with the electric-vector polarization angle rotating at a constant rate of $\sim 80 \thinspace \mathrm{deg} \thinspace \mathrm{day}^{-1}$. We test for variability in the Stokes parameters of \onees{} using a $\chi^{2}$-based hypothesis test. For this test, the null hypothesis is the assumption that the true Stokes parameters are equal to the mean value in each of $N=10$ time bins. These time bins are explicitly assigned to result in 4 time bins associated with the first segment and 6 with the second.  For this test, we use polarization cubes in the $2-6 \keV$ band in order to maximize the signal-to-noise of the polarization measurement in each time bin. Comparing the as-measured Stokes parameters in each time bin with the time-integrated mean, we find that the total $\chi^{2}$ of this model is $\chi^{2} = 14.30$ with 18 degrees of freedom. Similarly ``good'' values of $\chi^{2}$ are obtained for all values of $N$ in the range of $8-20$, suggesting that the lack of variability is not an artifact of our choice of the number of time bins. We therefore conclude there is insufficient evidence to reject a null hypothesis that the Stokes parameters in all of the time bins are statistically consistent with the time-integrated average. The variation in Stokes parameters with time are visualized in Figure \ref{fig:variability}. 

\section{Radio and Optical observations}\label{sec:multi_pol}
\begin{figure*}
    \centering
     \includegraphics[width=\textwidth]{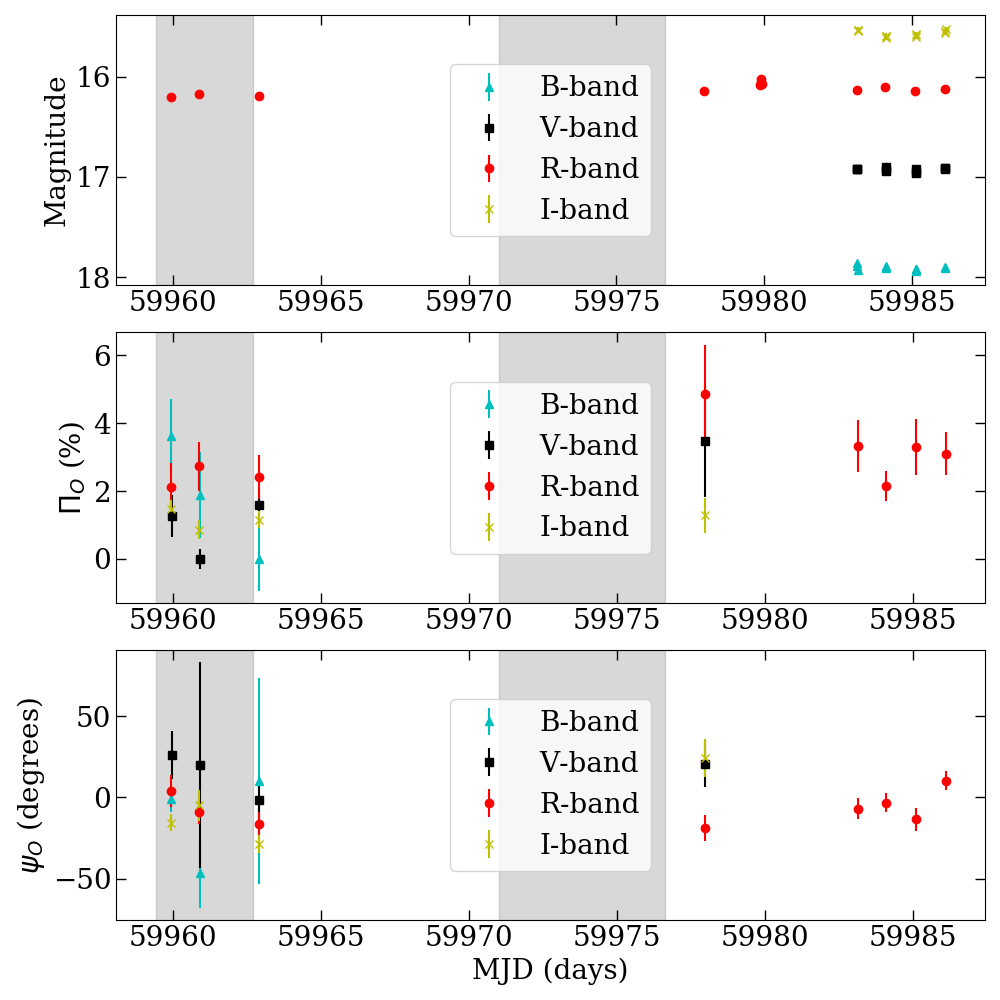}
    \caption{Multi-wavelength optical observations of \onees. The panels show the optical brightness (top), polarization degree (middle), and polarization angle (bottom). The IXPE observing periods are marked with the gray shaded regions. A correction for the host galaxy contribution has been applied exclusively to the R-band magnitudes and polarization measurements. }
    \label{plt:multi_obs}
\end{figure*}

During the IXPE observation we coordinated a radio, radio-millimeter (mm) and optical campaign with the Effelsberg 100-m radio telescope, the SubMillimeter Array (SMA),  the Nordic Optical Telescope (NOT), the Perkins telescope, and the Observatorio de Sierra Nevada (OSN). Observations with the Effelsberg and SMA  were obtained at 6 cm (4.85 GHz) and 1.3 mm (225.538 GHz) within the \emph{Monitoring the Stokes \textbf{Q}, \textbf{U}, \textbf{I} and \textbf{V} \textbf{E}mission of AGN jets in \textbf{R}adio} (QUIVER) program \citep{Kraus2003,Myserlis2018} and the \emph{\textbf{SMA} Monitoring of AGNs with \textbf{POL}arization} (SMAPOL) program \citep{Marrone2008}, respectively. Observations and data reduction at the NOT were performed with the  Alhambra Faint Object Spectrograph and Camera (ALFOSC) and the semi-automatic pipeline of Tuorla Observatory \citep{Hovatta2016,Nilsson2018}. Photometric observations from the Sierra Nevada observatories were also obtained in R-band and analysed using standard procedures. Additional R-band polarimetric observations and BVI photometry were obtained using the PRISM camera mounted on the 1.8 $\m$  Perkins Telescope following standard analysis procedures and differential photometry to estimate the brightness and polarization parameters. A more detailed description of the observations and data analysis for all the observatories used in this work can be found in \cite{Liodakis2022,DiGesu2022,Middei2023}, and \cite{Peirson2023}.

\onees{} has a bright host galaxy, the starlight from which dominates the emission at optical wavelengths. This results in the source appearing to be less variable and less polarized than is the case for the active nucleus alone. To correct for the host-galaxy contribution and estimate the intrinsic polarization degree ($\Pi_\mathrm{intr}$) of the source, we have performed detailed modeling of the host galaxy flux distribution following the method of \cite{Nilsson2007}. We estimate the contribution of the host galaxy to the R-band emission ($I_\mathrm{host}$) to be 0.54~mJy for an aperture with a 5$^{\prime\prime}$ radius and 0.67~mJy for 7.5$^{\prime\prime}$. We then use the estimated $I_\mathrm{host}$ for the apertures of the respective telescopes to correct the observed polarization degree $\Pi_\mathrm{obs}$. This is achieved by subtracting the host contribution from the total flux density ($I$), $\Pi_\mathrm{intr}= \Pi_\mathrm{obs}\times{I}/(I-I_\mathrm{host})$, following \cite{Hovatta2016}. Only the R-band $\Pi_\mathrm{O}$ estimates have been corrected. 

All of the multi-wavelength observations during the IXPE pointings are displayed in Fig. \ref{plt:multi_obs}. \onees{} is faint in radio (0.06 Jy at 4.85~GHz and 0.01 Jy at 225.5~GHz) which prevents us from detecting polarization at the 3$\sigma$ level. A QUIVER observation during the first segment of the IXPE observation yields an upper limit of $ 7\%$ (99\% confidence). Similarly, two separate SMAPOL observations after the second IXPE segment (MJD~59980, 59981) yield $<22\%$ (99\% confidence) and $<7\%$ (99\% confidence), respectively. For the NOT observations during the first IXPE segment, $\Pi_\mathrm{O}=2.42\pm0.72 \%$ along $\psi_\mathrm{O}=-2.4\degr \pm 8.5\degr$. We were unable to obtain contemporaneous optical observations during the second IXPE measurement. However, a few days later, the Perkins telescope measured $\Pi_\mathrm{O}=3.2\pm0.7 \%$ along a $\psi_\mathrm{O}=-5.1\degr \pm 8.7\degr $, consistent with the NOT results. This suggests similar levels of optical polarization for both IXPE observations.

\section{Discussion and Conclusions}\label{sec:disc_conc}

The \textit{IXPE} observations of \onees{} have detected X-ray polarization of $\Pi_\mathrm{X} \sim 17.9\%$ along $\psi_\mathrm{X} \sim 25^{\circ}$. The polarization degree we measure for this blazar is significantly higher than that measured during the first \textit{IXPE} observation of Mrk 501 \citep[$\sim 10\%$][]{Liodakis2022} and similar to that measured during the first \textit{IXPE} observation of Mrk 421 \citep[$\sim 15\%$;][]{DiGesu2022}. As for these two HSP blazars, the X-ray polarization degree of \onees{} is significantly higher than the optical values of $\sim 2\%$.  In this case, the ratio of X-ray to optical $\Pi$ is $>7$ making it the most strongly chromatic source in polarization observed thus far. 

\onees{} is not the first HSP blazar with higher degree of X-ray polarization than at longer wavelengths, yet with similar (within a few degrees) polarization angles in X-ray, optical, and radio bands. As discussed at length in \citet{DiGesu2022} and \citet{Liodakis2022}, the increasing polarization degree with energy suggests that the electrons generating these photons are accelerated at a shock. In this model, the X-ray photons are
generated by synchrotron radiation from electrons immediately downstream of the shock front, while lower energy photons, because of their longer radiative cooling time, are generated by electrons further downstream where the magnetic fields are more turbulent \citep{Marscher1985}. 

There remains one crucial difference between the results of \onees{} and those of Mrk 501 discussed in \cite{Liodakis2022}, however. In Mrk 501 the X-ray polarization angle was parallel to the position angle of the radio jet. In \onees{} the radio jet has an apparent position angle of $\sim 160^{\circ}$ east of north \citep{Piner2018} while the X-ray polarization angle is $\sim 30^{\circ}$, hence there is no obvious relationship between the polarization angle and the radio jet position angle. A similar mismatch was observed in the first observation of Mrk 421 \citep{DiGesu2022}, although later observations have also shown clear evidence of time variability in Mrk 421's polarization angle \citep{DiGesu2023}. The agreement between the X-ray polarization angle and the radio jet position angle in Mrk 501 was considered an important clue suggesting the presence of energy-stratified shock acceleration in the jet. It remains unclear how to reconcile the disagreement between the X-ray polarization angle and the radio jet direction for Mrk 421 and \onees{} with the predictions of an energy-stratified shock when all three blazars show the same wavelength-dependent polarization degree. However, we note that the estimates for the position angle of the jet are not contemporaneous. There is now a wealth of evidence showing that jets can change their position angle over time, such as in NRAO 150 \citep{Agudo2007}, OJ~287 \citep{Britzen2018},
PG1553+113 \citep{Lico2020}, and others \citep{Lister2013,Weaver2022}. In some cases, much larger position angle variations than the $50^{\circ}$ mismatch we observe have been seen over a few years \citep{Lister2013,Weaver2022}. Further monitoring of both the X-ray polarization and jet structure is needed to determine the extent to which such mismatches might be attributed to variability of the jet and polarization directions. The optical polarization angle of $\sim 0^{\circ}$ also appears to have no obvious relationship to either the X-ray polarization angle or the radio jet, although the polarization angles are closer to each other than to the radio jet's position angle. Fully accounting for the apparent discrepancy between the two polarization angles is beyond the scope of this paper, but does provide additional evidence that the regions where the optical and X-ray photons are generated appear to be largely disconnected from one another.

The reasonably high polarization degree, consistent with less extreme HSP blazars such as Mrk 501 and Mrk 421, further complicates any effort to account for the extreme properties of \onees{}. Recent single-zone \citep[e.g.][]{Tavecchio2022} and two-zone models \citep[e.g.][]{AguilarRuiz2022} often predict the electrons responsible for the X-rays originate in shock acceleration in highly turbulent regions of the magnetized jet plasma, where the magnetic fields are not expected to have any coherent direction. The relatively high polarization degree we observe is therefore in tension with such models. A possible way to reconcile these apparently disparate observations is to identify different length scales for the shock acceleration and synchrotron radiation - the magnetic fields may be turbulent on the small scales where electron acceleration occurs but more ordered and structured on the larger scales where these same high energy electrons are emitting their X-rays.  Other proposed models such as the multiple shock model proposed in e.g. \cite{Zech2021} do not explicitly require turbulence, but further testing and simulation work will be required to determine if such a model is able to reproduce the polarization results presented in this work. Magnetic reconnection \citep[e.g.][and references therein]{Matthews2020} appears to be disfavored as a viable particle acceleration model based on the low magnetic field strengths estimated from simple SSC model fits in \citet{Kaufmann2011} and \citet{Costamante2018}, but given the unusual fit parameters from these models there is reason to question whether or not this estimated magnetic field strength is an accurate measurement of what is physically present in the jet.  Although it is beyond the scope of this paper to fully develop a possible particle acceleration model that fully explains the polarization signal and multiwavelength properties of \onees{}, the fact that it has a similar X-ray polarization degree to Mrk 501 and Mrk 421 (despite the very different properties of these three blazars at $\mathrm{TeV}$ energies) is new evidence that informs future theoretical work understanding the most extreme blazar jets. Single zone acceleration models are more favored for less extreme HSP blazars due to correlations between the X-ray and gamma-ray light curves \citep[e.g.][and references therein]{Katarzynski2005}, but such a hypothesis has not been fully tested with \onees. The main reason this test has not yet been performed is that the light curves of \onees{} (in particular the gamma-ray light curve) show less variability than the less extreme blazars.  Further observations of \onees{} with X-ray and gamma-ray telescopes\footnote{In particular, future gamma-ray telescopes with higher sensitivity may be able to measure the variability of \onees{} on time scales that are unfeasible with current facilities. } including \ixpe{} will help identify the extent to which this source is variable in either flux or spectral properties. We find different spectral parameters than previous observations \citep[e.g.][]{Costamante2018} despite no obvious changes in its detected X-ray flux. It remains clear that the IXPE results for this extremely high synchrotron peaked blazar will require further modeling efforts to fully reconcile the extreme photon energies with the \ixpe{} measurements.  Simultaneous \ixpe{} and TeV observations may help further elucidate still unanswered questions about particle acceleration within the jet of this AGN.


\facilities{Imaging X-ray Polarimetry Explorer (\ixpe), XMM-Newton, Neil Gehrels Swift Observatory, Effelsberg 100\m{}  radio telescope, SubMillimeter Array, Nordic Optical Telescope, Perkins Telescope, Observatorio de Sierra Nevada} 

\software{\textsc{ixpeobssim} \citep{Baldini2022}, \textsc{XSPEC} \citep{Arnaud1996}, \textsc{XRTDAS} \footnote{\url{https://swift.gsfc.nasa.gov/analysis/xrt_swguide_v1_2.pdf}} \citep{Burrows2000},  \textsc{XMM-SAS} \citep{SAS2014} }

\section*{Acknowledgments}

 The Imaging X-ray Polarimetry Explorer (IXPE) is a joint US and Italian mission.  The US contribution is supported by the National Aeronautics and Space Administration (NASA) and led and managed by its Marshall Space Flight Center (MSFC), with industry partner Ball Aerospace (contract NNM15AA18C).  The Italian contribution is supported by the Italian Space Agency (Agenzia Spaziale Italiana, ASI) through contract ASI-OHBI-2017-12-I.0, agreements ASI-INAF-2017-12-H0 and ASI-INFN-2017.13-H0, and its Space Science Data Center (SSDC), and by the Istituto Nazionale di Astrofisica (INAF) and the Istituto Nazionale di Fisica Nucleare (INFN) in Italy.
This research used data products provided by the IXPE Team (MSFC, SSDC, INAF, and INFN) and distributed with additional software tools by the High-Energy Astrophysics Science Archive Research Center (HEASARC), at NASA Goddard Space Flight Center (GSFC).
  The IAA-CSIC co-authors acknowledge financial support from the Spanish "Ministerio de Ciencia e Innovaci\'{o}n" (MCIN/AEI/ 10.13039/501100011033) through the Center of Excellence Severo Ochoa award for the Instituto de Astrof\'{i}isica de Andaluc\'{i}a-CSIC (CEX2021-001131-S), and through grants PID2019-107847RB-C44 and PID2022-139117NB-C44. The POLAMI observations were carried out at the IRAM 30m Telescope. IRAM is supported by INSU/CNRS (France), MPG (Germany), and IGN (Spain).
  The Submillimetre Array is a joint project between the Smithsonian Astrophysical Observatory and the Academia Sinica Institute of Astronomy and Astrophysics and is funded by the Smithsonian Institution and the Academia Sinica. Mauna Kea, the location of the SMA, is a culturally important site for the indigenous Hawaiian people; we are privileged to study the cosmos from its summit. Some of the data reported here are based on observations made with the Nordic Optical Telescope, owned in collaboration with the University of Turku and Aarhus University, and operated jointly by Aarhus University, the University of Turku, and the University of Oslo, representing Denmark, Finland, and Norway, the University of Iceland and Stockholm University at the Observatorio del Roque de los Muchachos, La Palma, Spain, of the Instituto de Astrofisica de Canarias. E. L. was supported by Academy of Finland projects 317636 and 320045. The data presented here were obtained [in part] with ALFOSC, which is provided by the Instituto de Astrofisica de Andalucia (IAA) under a joint agreement with the University of Copenhagen and NOT. Part of the French contributions is supported by the Scientific Research National Center (CNRS) and the French spatial agency (CNES). The research at Boston University was supported in part by National Science Foundation grant AST-2108622, NASA Fermi Guest Investigator grants 80NSSC21K1917 and 80NSSC22K1571, and NASA Swift Guest Investigator grant 80NSSC22K0537. Some of the data are based on observations collected at the Observatorio de Sierra Nevada, owned and operated by the Instituto de Astrof\'{i}sica de Andaluc\'{i}a (IAA-CSIC). Further data are based on observations collected at the Centro Astron\'{o}mico Hispano en Andalucía (CAHA), operated jointly by Junta de Andaluc\'{i}a and Consejo Superior de Investigaciones Cient\'{i}ficas (IAA-CSIC). This work was supported by NSF grant AST-2109127. We acknowledge the use of public data from the Swift data archive. Based on observations obtained with XMM-Newton, an ESA science mission with instruments and contributions directly funded by ESA Member States and NASA. Partly based on observations with the 100-m telescope of the MPIfR (Max-Planck-Institut f\"ur Radioastronomie) at Effelsberg. Observations with the 100-m radio telescope at Effelsberg have received funding from the European Union’s Horizon 2020 research and innovation programme under grant agreement No 101004719 (ORP). I.L was supported by the NASA Postdoctoral Program at the Marshall Space Flight Center, administered by Oak Ridge Associated Universities under contract with NASA.



\bibliography{AllRefs}
\bibliographystyle{aasjournal}

\allauthors


\end{document}